# Electromagnetic stress tensor for amorphous metamaterial medium

Neng Wang,[1,2] Shubo Wang,[3] and Jack Ng[1,4*]

[1] *Department of Physics, Hong Kong Baptist University, Hong Kong, China*

[2] *Department of Physics and Institute for Advanced Study, The Hong Kong University of Science and Technology, Hong Kong, China*

[3] *Department of Physics, City University of Hong Kong, Hong Kong, China*

[4] *Institute of Computational and Theoretical Studies, Hong Kong Baptist University, Hong Kong, China*

[*]Correspondence to: jacktfng@hkbu.edu.hk

We analytically and numerically investigated the internal optical forces exerted by an electromagnetic wave inside an amorphous metamaterial medium. We derived, by using the principle of virtual work, the Helmholtz stress tensor, which takes into account the electrostiction effect. Several examples of amorphous media are considered, and different electromagnetic stress tensors, such as the Einstein-Laub tensor and Minkowski tensor, are also compared. It is concluded that the Helmholtz stress tensor is the appropriate tensor for such systems.

## I. Introduction

Optical trapping refers to the spatial confinement and manipulation of small particles by using optical forces [1–7]. Although optical trapping has been invented for over 40 years, development of its quantitative theory took place only in recent years [8-12]. However, a much less developed but yet important area of optical manipulation, namely optical stretching or deforming [13-15], is still not receiving sufficient attention. Here we address this issue by studying the internal optical forces acting on amorphous metamaterials. Such information would be useful, for example, in determining the deformation of a particle in an optical field.

It is well known that the total optical force acting on a scattering object can be calculated by integrating the appropriate electromagnetic stress tensor over a closed surface enclosing the object [16]. For typical background media for optical trapping, such as air and water, the appropriate tensors are the Maxwell stress tensor and the Minkowski tensor, respectively [17]. However, the determination of the force distribution inside or near the boundary of the scatter is far more challenging, because one will have to know the stress tensor inside the object. However, in the literature, there are multiple formulations of stress tensor and different experiments tend to support different stress tensors [31-37], which makes the issue even more confusing. There is no consensus on which one should be adopted inside the object itself [18-30].

To take the microscopic structure of the amorphous medium into account, we consider a model system consisting of a large number of discrete subwavelength elements. Such a system corresponds to those man-made materials that are well studied in the metamaterial community [38-43]. The problem can be addressed in either the microscopic or macroscopic levels. In the microscopic treatment, the scattering of light by the discrete elements are solved numerically to obtain the total microscopic field, which can then be used to calculate the optical force acting on each discrete elements using the Maxwell stress tensor (valid because the background is air). Since such an approach is rigorous, the results can be treated as the benchmark. However, the computation using the microscopic approach can be expensive, especially in 3D system. In the macroscopic treatment, we use the effective medium theory to model the discrete subwavelength elements and then solve for the macroscopic field, which can be used to calculate the optical force or stress by using a macroscopic stress tensor. We remark that in general, the macroscopic treatment is much more computational efficient.

The forces computed by different macroscopic stress tensors, such as the Einstein-Laub stress tensor (ELST), Minkowski stress tensor (MST), and Helmholtz stress tensor (HST), are compared with the benchmarking results to check their accuracies. Based on this idea, recently Sun et al [44] and Wang et al [45] showed that both the body and boundary forces for such a material depend not only on the macroscopic parameters (i.e. the numerical values of the effective permittivity and permeability), but also the microscopic lattice structure. They found that both the body and boundary forces can be predicted by using the HST. And more recently, Wang et al [46] extended the HST for the bi-anisotropic metamaterials. The electrostrictive and magnetostrictive terms in the HST, which capture the information of the microscopic lattice structure, such as lattice type and filing ratio, play important roles in the body and boundary force calculations. Other stress tensors such as the Maxwell stress tensor, MST, and ELST, which lack the electrostrictive and magnetostrictive terms, missed the information of microscopic lattice structure, and therefore cannot describe the body and boundary forces correctly.

The aim of this paper is two folds. First, we unveil why the HST can correctly captures the details of the microstructure. We show that it is due to the equality of the average energy in the discrete medium and effective continuous medium over a large length scale [47], and also the spatial average relation between the microscopic and macroscopic fields in the long wavelength limit [48]. Second, we compare the HST, ELST and MST in calculating optical forces inside a 2D amorphous medium. The difference between the HST and ELST is $\varepsilon_0(\varepsilon_e - 1)^2 |\mathbf{E}|^2 /4$. Whenever this term dominates, the HST outperforms the ELST.

While in some lossy cases where the difference term is small, HST produces roughly the same results as ELST. The reason we focus on amorphous system is that the electrostrictive and magnetostrictive terms are single Bloch $k$-dependent for regular lattices (e.g. square lattice) [36], which is ill-defined for lossy cases. However, this is not an issue for amorphous systems where the electrostrictive and magnetostrictive terms are $k$-independent.

In Sec. II, we outlined the derivation of the HST for amorphous discrete system, and explain the equivalence of the free energy in both the discrete system and effective medium. This lays the foundation for HST. Then, in Sec. III, we numerically verified the analytical predictions and compare the forces determined by various stress tensors for both lossless and lossy amorphous systems. Finally, we conclude in Sec. IV.

## II. Derivation of Helmholtz stress tensor for amorphous medium

For consistency and completeness, the HST is first derived from the principle of virtual work in the quasi-static limit, although similar derivation can be found in ref. [49]. Then we show that the HST can also be applied for man-made metamaterials consisting of subwavelength elements.

Consider a small rectangular area $dS = ab$ inside an 2D amorphous medium with $a \ll \lambda$ and $b \ll \lambda$, as shown in Fig.1. Since the electric field inside $dS$ is nearly uniform, the time-averaged total electric energy of this area is given by $-1/4\,\text{Re}(\mathbf{E}\cdot\mathbf{D}^*)dS$. If we subject one of the boundaries, i.e. such as the right boundary, a virtual infinitesimal displacement $\xi$, then the change in the total electric energy $W_e$ for this area is just the work done by the electric component of the boundary force $T_{ik,e}\xi_i n_k b$, where $T_{ik,e}$ is the electric component of stress tensor and **n** is the unit normal vector of the boundary. Hence, we have

$$\delta W_e = T_{ik,e}\xi_i n_k b = -\frac{1}{4}\text{Re}\{\mathbf{E}\bullet\mathbf{D}^*\}b\mathbf{n}\bullet\boldsymbol{\xi} + \frac{\partial W_e}{\partial \mathbf{E}}\bullet\delta\mathbf{E}^* + \frac{\partial W_e}{\partial \text{Re}\{\varepsilon_e\}}\frac{\partial \text{Re}\{\varepsilon_e\}}{\partial \rho}\delta\rho. \quad (1)$$

The first term on the right hand side is due to the change in total area, while the second and third terms are due to the change in electric field and permittivity, respectively. Here the permittivity of the material is assumed to be depending only on the mass density $\rho$, which is true for an amorphous medium. For isotropic materials where $\mathbf{D} = \varepsilon_e \mathbf{E}$, the second term can be re-written as

$$\frac{\partial W_e}{\partial \mathbf{E}}\bullet\delta\mathbf{E}^* = -\frac{1}{4}\text{Re}\{\mathbf{D}^*\bullet\delta\mathbf{E} + \mathbf{E}\bullet\delta\mathbf{D}^*\}dS = -\frac{1}{2}\text{Re}\{\mathbf{D}^*\bullet\delta\mathbf{E}\}ab \quad (2)$$

Note that the potential of each point on the boundary of $dS$ remains invariant during the deformation [49], namely $\mathbf{E'}\bullet\mathbf{n}a + \mathbf{E'}\bullet\boldsymbol{\xi} = \mathbf{E}\bullet\mathbf{n}a$ and $\mathbf{E'}\times\mathbf{n}b = \mathbf{E}\times\mathbf{n}b$, we thus have

$$\delta\mathbf{E} = \mathbf{E'} - \mathbf{E} = -\mathbf{n}\frac{\mathbf{E}\bullet\boldsymbol{\xi}}{a} . \quad (3)$$

For the third term, the conservation of the total mass implies $\delta(\rho dS) = 0$, one thus has

$$\delta\rho = -\frac{\rho\delta(dS)}{dS} = -\frac{\rho\mathbf{n}\bullet\boldsymbol{\xi}}{a} . \quad (4)$$

As a result, the third term becomes

$$\frac{\partial W_e}{\partial \text{Re}(\varepsilon)}\frac{\partial \text{Re}(\varepsilon)}{\partial \rho}\delta\rho = \frac{1}{4}(\mathbf{E}\bullet\mathbf{E}^*)dS\frac{\rho\mathbf{n}\bullet\boldsymbol{\xi}}{a}\frac{\partial \text{Re}(\varepsilon)}{\partial \rho} = \frac{1}{4}|\mathbf{E}|^2 \mathbf{n}\bullet\boldsymbol{\xi}b\rho\frac{\partial \text{Re}(\varepsilon)}{\partial \rho} . \quad (5)$$

Substituting Eqs.(2), (3) and (5) into Eq.(1), one arrives at

$$T_{ik,e} = -\frac{1}{4}\text{Re}(\mathbf{E} \cdot \mathbf{D}^*)\delta_{ik} + \frac{1}{2}\text{Re}(E_i D_k^*) + \frac{1}{4}|\mathbf{E}|^2 \frac{\partial \text{Re}(\varepsilon)}{\partial \rho}\rho\delta_{ik} , \qquad (6)$$

where $\delta_{ik}$ is the Kronecker delta function. For the sake of mathematical simplicity, here we take the relative permeability of the material to be $\mu_r = 1$. Accordingly, one has

$$T_{ik,helm} = \frac{1}{2}\text{Re}\{E_i D_k^* - \frac{1}{2}(\varepsilon - \rho\frac{\partial \varepsilon}{\partial \rho})\delta_{ik}|\mathbf{E}|^2 + \mu_0 H_i H_k^* - \frac{1}{2}\mu_0 \delta_{ik}|\mathbf{H}|^2\} , \qquad (7)$$

Eq. (7) can be extended straightforwardly to magnetic materials by including the analogous magnetostrictive terms [37]. We note that the HST reduces to the Maxwell stress tensor for E-z polarization [36], therefore we will focus on the H-z polarization.

The HST is valid for metamaterials only if the electric energy density for the discrete amorphous medium equals to that of the effective medium on average. However, this is not necessarily true for a discrete system where one deals with the microscopic fields, such as the amorphous medium shown in Fig. 2, which is a domain consists of many subwavelength sized cylinders randomly distributed inside. For separations between adjacent cylinders small compared to the incident wavelength, this domain can be treated as a continuous effective medium [Fig. 2(b)], with effective relative permittivity $\varepsilon_e$. Apparently, the electromagnetic fields inside these two media (see Fig. 2(a) and (b)) are different, even though their scattering properties are very similar. Clearly, the equality of the electromagnetic energy for the two media, nor the applicability of the HST to effective medium (b), is obvious. Nevertheless, we analytically showed (see Appendix) that for an area enclosing sufficiently many small cylinders, the total electric energy of the discrete medium is indeed equal to that of the corresponding effective continuous medium [47,48]. Similarly, one can show the same for the magnetic energy as well.

In the long wavelength limit, the effective permittivity needed in Eq. (7) is given by the Clausius-Mossotti relation, and for H-z polarization, it takes the form $\varepsilon_e = (1+pM)/(1-pM)$, where $M = (\varepsilon_c - 1)/(\varepsilon_c + 1)$, $\varepsilon_c$ is the relative permittivity of the cylinders and $p$ is the filling ratio. Accordingly, the electrostrictive term becomes

$$-\rho \frac{\partial \varepsilon_e}{\partial \rho} |\mathbf{E}|^2 = -p \frac{\partial \varepsilon_e}{\partial p} |\mathbf{E}|^2 = -\frac{\varepsilon_e^2 - 1}{2} |\mathbf{E}|^2 \ . \qquad (11)$$

### III. Numerical results and discussions

We construct the 2D discrete amorphous medium by randomly locating 6050 small identical nonmagnetic dielectric cylinders within a circular domain of radius $r_0 = \lambda$, see Fig. 2(a).

The fact that the small cylinders are placed randomly makes the boundary deviates slightly from a perfect sphere. To fix this problem, we defined an effective radius for the effective medium, which is slightly smaller than $r_0 = \lambda$. By averaging the position fluctuation in the boundary (see Appendix for details), the effective radius of the effective medium is determined to be $r'_0 = 0.988\lambda$, as shown in Fig. 2(b). The radius of each small cylinder is $r_2 = 3 \times 10^{-3} \lambda$ and the filling ratio is $p = 0.0558$. The relative effective permittivity $\varepsilon_e$ of the medium can be tuned by varying the relative permittivity of the small cylinders. The incident plane wave propagates from left to right, and exerts an optical force on each small cylinder. These forces can be rigorously calculated by the using the generalized Lorenz-Mie theory and the Maxwell stress tensor (valid because the integration is to be performed over air), where the scattering coefficients for each cylinder are determined by the multiple scattering theory [50-53]. Due to the random fluctuations of the field inside the amorphous medium, comparing the force acting on a signle cylinder element is not meaningful. We thus focus on the total force acting on a finite area enclosed by the dashed line in Fig. 2(a), which can be obtained microscopically by summing the forces acting on each consisting cylinder element. On the other hand, the optical force acting on the same region inside the effective medium (b) can also be calculated macroscopically using HST. The two can then be compared to verify our theory. For comparison, we will also consider the MST and ELST, which are frequently considered in previous studies [24-25, 27-29]. The MST is expressed as

$$T_{ik,\min} = \frac{1}{2}\mathrm{Re}\{E_i D_k^* - \frac{1}{2}\varepsilon\delta_{ik}|\mathbf{E}|^2 + \mu_0 H_i H_k^* - \frac{1}{2}\mu_0\delta_{ik}|\mathbf{H}|^2\},$$

while the ELST is

$$T_{ik,eins} = \frac{1}{2}\mathrm{Re}\{E_i D_k^* - \frac{1}{2}\varepsilon_0\delta_{ik}|\mathbf{E}|^2 + \mu_0 H_i H_k^* - \frac{1}{2}\mu_0\delta_{ik}|\mathbf{H}|^2\}.$$

Consider the lossless systems shown in Fig. 3 (a), (b) and (c) where the total optical force acting on the concentric circular regions of radius $r = 0.4\lambda, 0.55\lambda, 0.8\lambda$ are plotted, respectively. For each value of $\varepsilon_e$, 40 samples of discrete amorphous medium are numerically calculated and the result for each sample is denoted by a blue circle in Fig. 3, while the ensemble averages are given by the blue dashed lines. The results calculated by the HST and the ELST are represented by the red and green solid lines, respectively. Note that for the lossless system, the internal optical force calculated by the MST is exactly zero, therefore they are not shown here. The total electric energies inside these regions are shown in Fig. 3(d), (e), and (f), where the blue circles are for the discrete medium while the red lines are for the effective continuous medium. We conclude that although the electric field distribution inside the discrete and effective continuous mediums are very different, the total electric energy is almost the same, in agreement with our analytical results, see Appendix A.

For small domain not enclosing sufficiently many small cylinders, as in the case of $r = 0.4\lambda$, the standard deviations of the 40 samples are large such that their average may not be the true average and thus deviates from the macroscopic calculation by stress tensors. For $r = 0.8\lambda$, although sufficiently many cylinders are enclosed and the standard deviation is much smaller, the boundary of the concerned region is close to the boundary of the entire amorphous medium, where the effective medium description fails. Accordingly, the HST also fails to predict the average values of the discrete medium. A very good agreement is achieved at $r = 0.55\lambda$ where the concerned region is sufficiently large and far away from the boundary. The ELST deviates significantly from the benchmarking results, especially for large effective permittivity. The difference between the HST and ELST is $\varepsilon_0 (\varepsilon_e - 1)^2 |\mathbf{E}|^2 / 4$, which approaches zero when $\varepsilon_e$ approaches to 1, but deviates from zero when $\varepsilon_e$ is large. This explains why the ELST fails at large $\varepsilon_e$.

The effect of the domain size $r$ is studied in Fig. 4(a), where $\varpi$ is defined to be a measure of the relative difference in force averaged over $N$ spatial points.

$$\varpi = \frac{1}{N} \sum_{i=1}^{N} \frac{f_e^i - \overline{f}_d^i}{\overline{f}_d^i}, \qquad (12)$$

where $f_e^i$ is the forces given by one of the macroscopic stress tensor for a spatial point labeled by $i$, $\overline{f}_d^i$ is ensemble average of the random samples, and $N = 7$ is the total number of spatial points considered. Figure 4(a) shows that both the HST and the ELST work best when $0.5\lambda \leq r \leq 0.6\lambda$, due to the reasons explained above. However, the HST is much more accurate than the others, with a small relative difference of ~5%. It is expected that for small $r$, the HST should have correctly predicted the ensemble average. It is just that the 40 samples ensemble is too small to yield statistically significant results as fluctuation dominates. Figure 4(b) shows the relative standard deviation $\eta$ for the 40 samples as a function of $r$. As expected, the standard deviation gradually approaches zero when $r$ increases. We remark that the trend in accuracies of HST with respect to $r$ is typical that of effective medium, which is accurate for region large compare to the wavelength.

The situation for lossy system is similar, as shown in Fig. 5, when $\varepsilon_e = \varepsilon_{er} + i\varepsilon_{ei}$, and $\varepsilon_{ei} = 0.05$. As shown in Fig. 5 (d)-(f), even in the lossy case, the electric energy of the amorphous medium and effective medium are the same. For lossy systems, the MST gives nonzero forces inside the amorphous medium, which is plotted by the black solid lines in Fig 5(a)-(c). Nevertheless, the prediction of the MST deviates significantly from that of the amorphous medium, due to the omission of the electrostriction term. For the HST, the best agreement is still achieved at around $0.5\lambda \leq r \leq 0.6\lambda$. $\varpi$ and $\eta$ as functions of $r$ are plotted in Fig 6. Similar to the lossless system, the HST outperforms others for all values of $r$, and the standard deviation approaches zero when $r$ increases.

We also studied the effect of the size of the amorphous medium. We reduce the incident wavelength by half while keeping all other parameters unchanged. In Fig. 7(a), we show $\varpi$ as a function of *r* for the lossless case. The HST is still much better than the ELST. The force versus the effective permittivity for the specific case of $r = 1.6\,\lambda$ marked by the black rectangle in Fig. 7(a) is shown in Fig. 7(b). It is clear that the HST agrees with the discrete amorphous medium well, while the ELST shows significant deviation.

Figure 8 shows the difference between the HST and ELST in the presence of losses. The discrepancy becomes small due to two reasons. First, the difference between the HST and ELST becomes smaller where there is a non-zero imaginary part in the effective permittivity. Second, the contribution of the common term, which is just the MST, becomes dominant, as shown by the black solid line in Fig. 8(b).

Finally, we remark that although our theoretical work indicates the HST should work for arbitrary effective permittivity, the numerical results are less accurate for high effective permittivity or loss, see Figs. 3, 5, 7 and 8. The underlining reasons are still under investigation.

## IV. Conclusions

In summary, we show analytically that the HST can be applied to calculate the optical force inside an amorphous metamaterial medium consisting of subwavelength elements. This is enabled by the equal-energy relationship between the amorphous medium and the effective medium, as well as the spatial average relation between the microscopic and macroscopic fields.

Using explicit numerical examples, we studied the performance of HST, ELST, and MST, for two-dimensional man-made amorphous medium. It is found that the electrostriction effect arising from the microscopic structures plays a very important role in optical forces, and that the HST is most accurate among those we considered. For lossy medium, the HST and ELST both tend to approach the MST and therefore all of them predict very similar forces.

## V. Acknowledgement

This work is supported by Hong Kong Research Grant Council grants AoE/P-02/12 and HKBU 209913. We thank Prof. Z. Q. Zhang and Dr. Kun Ding for useful comments and suggestions.

## Appendix A: The electric energy of a typical unit cell in amorphous discrete medium

Based on the coherent potential approximation (CPA) method [47,54,55], a typical unit cell of our circular amorphous discrete medium can be treated as a coated cylinder with $r_1, r_2$ being the outer and inner radii respectively, and $\varepsilon_c$ being the relative permittivity of the core, as shown in Fig. 2 (c). For H-z polarization, the effective permittivity of the medium is $\varepsilon_e = (1+pM)/(1-pM)$, where $M = (\varepsilon_c - 1)/(\varepsilon_c + 1)$, and $p = r_2^2 / r_1^2$ is the filling ratio. If we place the coated cylinder inside a continuous medium with relative permittivity $\varepsilon_e$, then since the permittivity is uniform, the scattering of the coated cylinder will vanish in the long wavelength limit. According to the Mie theory, the incident, scattered, and internal electromagnetic fields in the shell and core can be written as [52]

$$\mathbf{E}_i = \sum_n E_n p_n \mathbf{M}_n^{(1)}(k,\mathbf{r}), \quad \mathbf{E}_s = -\sum_n E_n a_n \mathbf{M}_n^{(3)}(k,\mathbf{r}), \tag{A1a}$$

$$\mathbf{E}_{ins}^1 = \sum_n E_n [f_n \mathbf{M}_n^{(1)}(k_1,\mathbf{r}) + g_n \mathbf{M}_n^{(3)}(k_1,\mathbf{r})], \quad \mathbf{E}_{ins}^2 = \sum_n E_n c_n \mathbf{M}_n^{(1)}(k_2,\mathbf{r}), \tag{A1b}$$

$$\mathbf{H}_i = -\frac{ik}{\omega}\sum_n E_n p_n \mathbf{N}_n^{(1)}(k,\mathbf{r}), \quad \mathbf{H}_s = \frac{ik}{\omega}\sum_n E_n a_n \mathbf{N}_n^{(3)}(k,\mathbf{r}), \tag{A1a}$$

$$\mathbf{H}_{ins}^1 = -\frac{ik_1}{\omega}\sum_n E_n [f_n \mathbf{N}_n^{(1)}(k_1,\mathbf{r}) + g_n \mathbf{N}_n^{(3)}(k_1,\mathbf{r})], \quad \mathbf{H}_{ins}^2 = -\frac{ik_2}{\omega}\sum_n E_n c_n \mathbf{N}_n^{(1)}(k_2,\mathbf{r}), \tag{A1b}$$

where

$$\mathbf{M}_n^{(J)} = [\frac{in}{\rho}z_n^{(J)}(\rho)\mathbf{e}_r - z_n'^{(J)}(\rho)\mathbf{e}_\phi]e^{in\phi}, \mathbf{N}_n^{(J)} = z_n^{(J)}(\rho)e^{in\varphi}\mathbf{e}_z, E_n = i^n, z_n^{(1)}(\rho) = J_n(\rho), z_n^{(3)} = H_n^1(\rho)$$

and $k = \sqrt{\varepsilon_e}k_0, k_1 = k_0, k_2 = \sqrt{\varepsilon_c}k_0$ with $k_0$ being the wave number in the vacuum. According to the standard electromagnetic boundary conditions, the tangential components of the electromagnetic fields are continuous, one could derive the expansion coefficients as for the fields

$$A_n = \frac{g_n}{f_n} = \frac{m_1 J_n(m_1 x) J'_n(m_2 x) - m_2 J'_n(m_1 x) J_n(m_2 x)}{m_2 H'_n(m_1 x) J_n(m_2 x) - m_1 H_n(m_1 x) J'_n(m_2 x)}, \tag{A2a}$$

$$\frac{f_n}{p_n} = \frac{J_n(y) H'_n(y) - J'_n(y) H_n(y)}{m_1[J(m_1 y) + A_n H_n(m_1 y)] H_n'(y) - [J'_n(m_1 y) + A_n H'_n(m_1 y)] H_n(y)}, \tag{A2b}$$

$$\frac{c_n}{p_n} = \frac{J'_n(m_1 x) + A_n H'_n(m_1 x)}{J'_n(m_2 x)} \frac{f_n}{p_n}, \tag{A2c}$$

where $x = ka, y = kb, m_1 = k_1/k, m_2 = k_2/k$. In the long wavelength limit $\rho \to 0$, according to Eq. (A2), all terms with $n \neq \pm 1$ will vanish. Using the asymptotic approximations for $\rho \to 0$:

$J_1(\rho) \sim \rho/2, J'_1(\rho) \sim 1/2, H_1(\rho) \sim -2i/(\pi\rho), H'_1(\rho) \sim 2i/(\pi\rho^2)$, one arrives at

$$\frac{f_1}{p_1} = \frac{f_{-1}}{p_{-1}} = \frac{2(\varepsilon_1 + \varepsilon_2)}{\varepsilon_1^2(1-p) + \varepsilon_2(1-p) + \varepsilon_1(1+\varepsilon_2)(1+p)} = \frac{1}{1-pM}, \quad (A3a)$$

$$\frac{g_1}{p_1} = \frac{g_{-1}}{p_{-1}} = -\frac{i\varepsilon_1(\varepsilon_1 - \varepsilon_2)}{\varepsilon_1^2(1-p) + \varepsilon_2(1-p) + \varepsilon_1(1+\varepsilon_2)(1+p)} \frac{\pi x^2}{2} = \frac{g'_1}{p_1} \frac{i\pi x^2}{4}, \quad (A3b)$$

$$\frac{c_1}{p_1} = \frac{c_{-1}}{p_{-1}} = \frac{4\varepsilon_1}{\varepsilon_1^2(1-p) + \varepsilon_2(1-p) + \varepsilon_1(1+\varepsilon_2)(1+p)} = \frac{2}{1+\varepsilon_c + p(1-\varepsilon_c)}, \quad (A3c)$$

$$\mathbf{M}_1^{(1)} = \frac{1}{2}(i\mathbf{e}_r - \mathbf{e}_\phi)e^{i\phi}, \quad \mathbf{M}_{-1}^{(1)} = \frac{1}{2}[i\mathbf{e}_r + \mathbf{e}_\phi]e^{-i\phi}, \quad \mathbf{M}_{n\neq\pm1}^{(1)} = 0, \quad (A3d)$$

$$\mathbf{M}_1^{(3)} = \frac{2i}{\pi\rho^2}(-i\mathbf{e}_r - \mathbf{e}_\phi)e^{i\phi}, \quad \mathbf{M}_{-1}^{(3)} = \frac{2i}{\pi\rho^2}[-i\mathbf{e}_r + \mathbf{e}_\phi]e^{-i\phi}, \quad \mathbf{M}_{n\neq\pm1}^{(3)} = 0, \quad (A3e)$$

Where $\varepsilon_1 = 1/\varepsilon_e$, $\varepsilon_2 = \varepsilon_c/\varepsilon_e$. Consequently, the incident electric field and electric fields in the shell and core are given by

$$\mathbf{E}_i = \frac{1}{2}[(ie^{i\phi}p_1 + ie^{-i\phi}p_{-1})\mathbf{e}_r + (-e^{i\phi}p_1 + e^{-i\phi}p_{-1})\mathbf{e}_\phi] = E_r\mathbf{e}_r + E_\phi\mathbf{e}_\phi, \quad (A4a)$$

$$\mathbf{E}_{ins}^1 = \frac{f_1}{p_1}(E_r\mathbf{e}_r + E_\phi\mathbf{e}_\phi) + \frac{g'_1}{p_1}\frac{x^2}{\rho^2}(E_r\mathbf{e}_r - E_\phi\mathbf{e}_\phi), \quad (A4b)$$

$$\mathbf{E}_{ins}^2 = \frac{c_1}{p_1}(E_r\mathbf{e}_r + E_\phi\mathbf{e}_\phi), \quad (A4c)$$

where $E_r$ and $E_\phi$ are the radial and azimuthal components of incident electric field. Then the total energy in the cell can be derived:

$$\begin{aligned}
W_e &= -\frac{1}{4}\text{Re}(\int_{shell}|\mathbf{E}_{ins}^1|^2 dS + \int_{core}\varepsilon_c|\mathbf{E}_{ins}^2|^2 dS \\
&= -\frac{1}{4}|\mathbf{E}_i|^2 \{\text{Re}(\varepsilon_c)|\frac{c_1}{p_1}|^2|\mathbf{E}_i|^2 \pi r_2^2 + 2\pi\int_{r_2}^{r_1}(|\frac{f_1}{p_1}|^2 r + |\frac{g'_1\varepsilon_e}{p_1}|^2 \frac{r_2^4}{r^3})dr\} \\
&= -\frac{1}{4}|\mathbf{E}_i|^2 \pi r_1^2 \text{Re}[\varepsilon_c |\frac{c_1}{p_1}|^2 p + |\frac{f_1}{p_1}|^2(1-p) + |\frac{g'_1\varepsilon_e}{p_1}|^2 p(1-p)] \\
&= -\frac{1}{4}|\mathbf{E}_i|^2 \pi r_1^2 \text{Re}(\varepsilon_e)
\end{aligned} \quad (A5)$$

Clearly the total electric energy of the unit cell given in (A5), which is for a coated cylinder and equivalent to the discrete amorphous medium, is just equal to what one would have when one replaces the coated cylinder by a homogenous cylinder with relative permittivity $\varepsilon_e$. We remark that similar derivation can be performed for the magnetic energy as well.

**Appendix B: Determining the effective size of the effective medium**

A portion of the circular amorphous medium near the boundary consisting of subwavelength dielectric cylinders is shown in Fig. 9. The orange circles denote the small cylinders which are randomly distributed inside a circular boundary. Part of the circular boundary of the medium is shown by the blue dashed arc line. Due to the fluctuations in radial distances between the outermost cylinders and the center of the circular region, the boundary of the effective medium is really not the perfect circle defined by the blue lines. Here, we define the radius of the effective medium to be the average radial distance for each outermost cylinders from the center of the circular region. The outermost cylinders are the ones inside the outermost layer which is marked by the red and blue dashed arc lines, see Fig. 9. However, the thickness of the outermost layer is not well defined. Here we choose the thickness of the layer to be the lattice constant in a square lattice with the same filling ratio, namely $d = \sqrt{\pi/p}\, r_c = \sqrt{\pi/N}\, r_0$, where $N$ is the total number of the cylinders and $r_0$ is the radius of the circular region. This is how we determine the radius of the effective medium to be $0.988\lambda$, which serves as a minor correction.

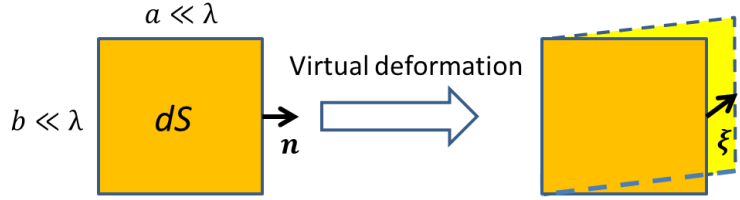

**Figure 1. The principle of the virtual work.** For a small area *dS*, the total electric energy is $-1/4\,\mathrm{Re}(\mathbf{E}\cdot\mathbf{D})dS$. If one boundary is subjected to a virtual infinitesimal displacement **ξ**, then the change in the total electric energy is equal to the work done by the electric component of the boundary force $T_{ik,e}\xi_i n_k b$, where $T_{ik,e}$ is the electric component of surface stress tensor and **n** is the unit normal vector of the boundary.

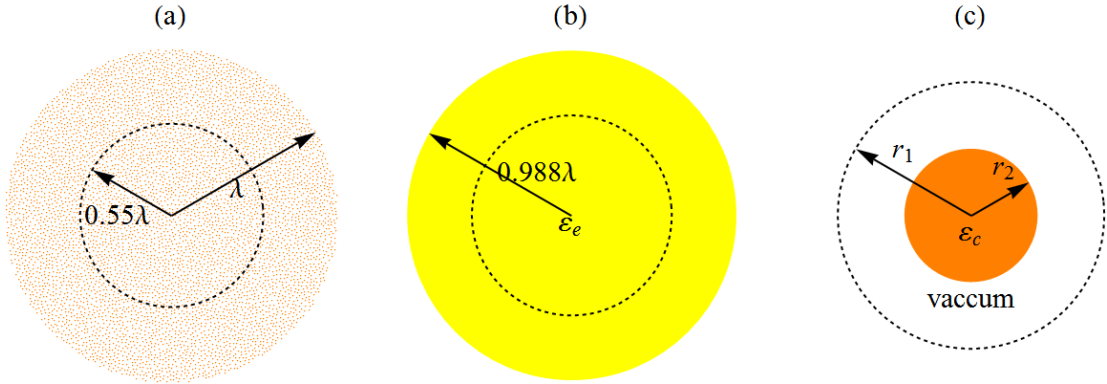

**Figure 2. Geometry of the problem.** (a) is a circular sample of an amorphous effective medium with radius λ and 6050 small identical cylinders randomly distributed inside. In the long wavelength limit, the domain can be treated as a continuous medium, with relative permittivity $\varepsilon_e$ and an effective radius of 0.988 λ (see appendix B for details), as shown in (b). Under light illumination, all cylinders are subjected to an optical force, which can be calculated individually by using the generalized Lorenz-Mie theory and Maxwell stress tensor. For the circular region enclosed by the black dashed circle of radius $r = 0.55$ λ, the total optical force is the sum of the optical forces acting on all small cylinders within. For the effective continuous medium, the optical force acting on the same region can be calculated by the Helmholtz, Einstein-Laub, and Minkowski stress tensors. (c) is a sketch of a typical unit cell.

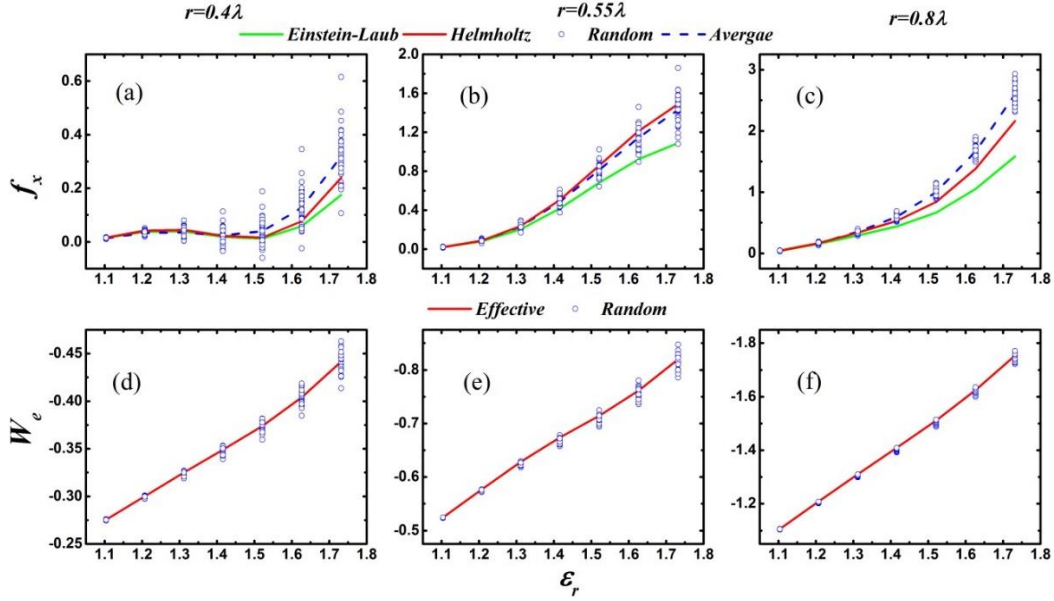

**Figure 3. Total optical forces and electric energies for lossless systems with different radius.** The total optical forces for radii $r = 0.4 \lambda$, $0.6 \lambda$, and $0.8 \lambda$, as functions of the effective relative permittivity $\varepsilon_e$, are shown in (a), (b), and (c), respectively. 40 samples of discrete system are generated and the forces for each sample are shown as a blue circle. The average of the 40 samples are plotted by the blue dashed lines. For the effective continuous medium, the optical force acting on the same regions are calculated by the HST and ELST as shown by the red and green solid lines, respectively. The total electric energies for the regions in the discrete (blue circles) and effective continuous (red solid lines) mediums are also plotted in (d), (e) and (f).

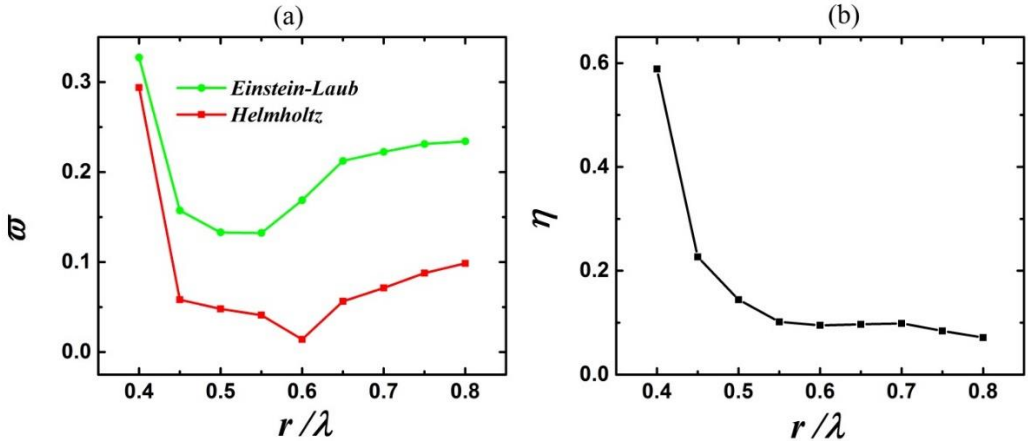

**Figure 4. $\varpi$ and $\eta$ as functions of the radius of the calculation region for lossless systems.** (a) $\varpi$ as a function of the radius of the calculation region for the HST and ELST are shown by the red and green lines, respectively. (b) The total relative standard deviations $\eta$ of the 40 rigorous results for regions of different sizes.

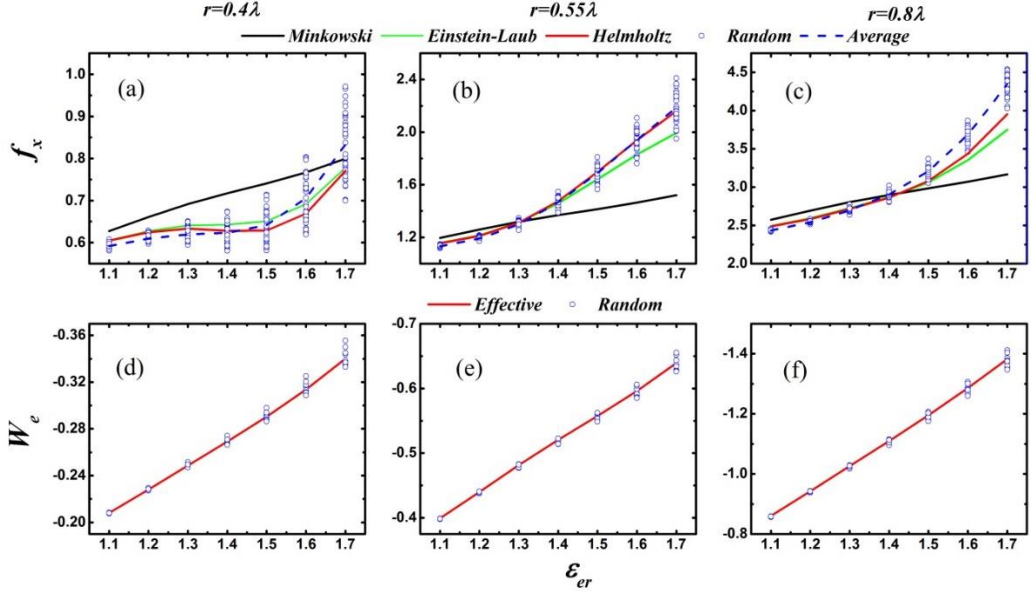

**Figure 5. Total optical forces and electric energies for lossy systems with different radius.** The total optical forces for radii $r = 0.4\lambda$, $r = 0.55\lambda$ and $r = 0.8\lambda$, as functions of real part of effective relative permittivity $\varepsilon_{er}$, are shown in (a), (b) and (c), respectively. The imaginary part of the effective permittivity is fixed at $\varepsilon_{ei} = 0.05i$. 40 samples of discrete medium are generated and the forces of each sample are shown by the blue circles. The averages of the 40 samples are denoted by the blue dashed line. For the effective continuous medium, the optical force acting on the same regions are calculated by the HST, ELST, and MST as shown by the red, green and black solid lines, respectively. The total electric energies for the regions in the discrete (blue circles) and effective continuous (red solid lines) mediums are shown in (d), (e) and (f).

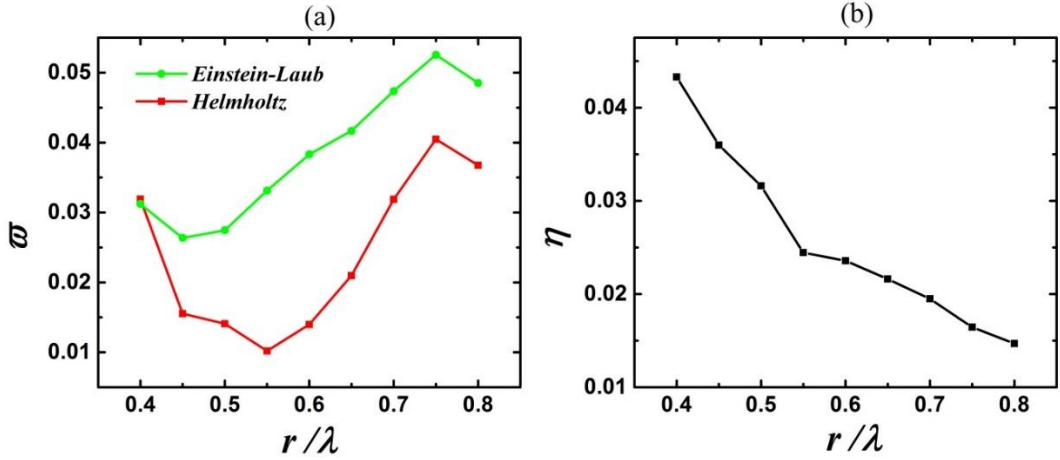

**Figure 6. ϖ and η as functions of the region radius for lossy system.** (a) ϖ for HST and ELST are plotted by the red and green lines, respectively. (b) η, the total relative standard deviations for the 40 samples.

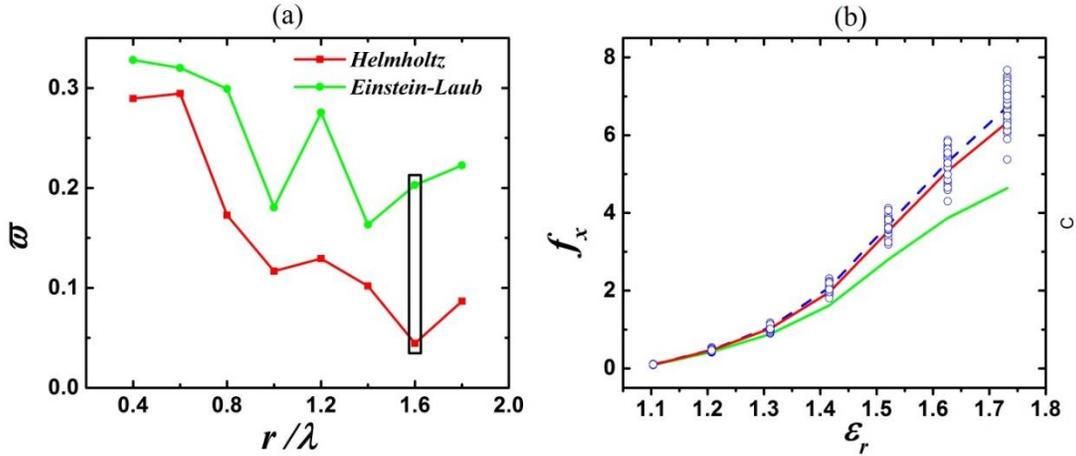

**Figure 7. Double sized lossless system.** (a) ϖ for HST and ELST are plotted by the red and green lines, respectively. (b) The total optical force for $r = 1.6\,\lambda$ [marked by the black rectangle in (a)] calculated by different stress tensors. The radii for the small cylinders and the entire circular domain are $6 \times 10^{-3}$ λ and 1.976 λ, respectively.

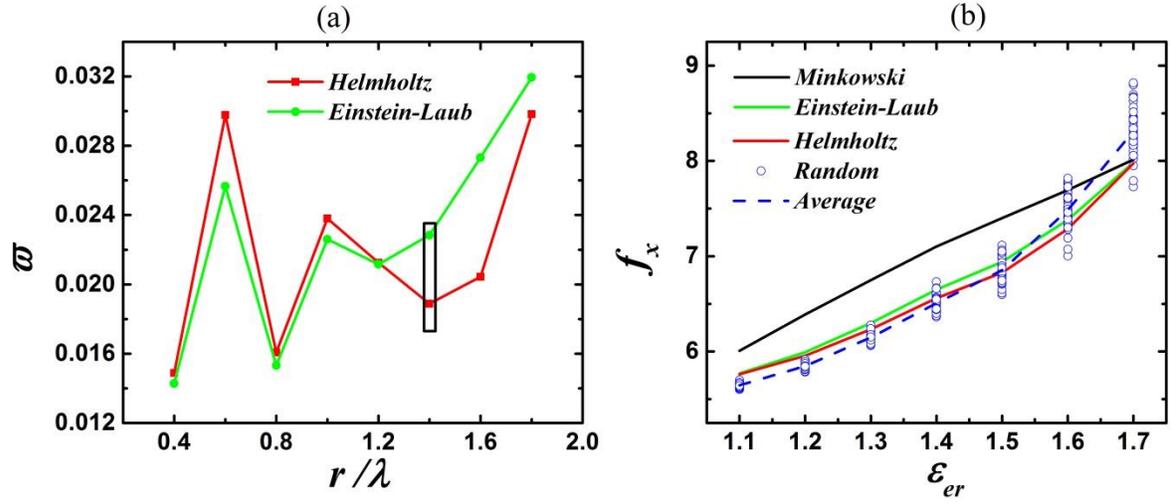

**Figure 8. Double-sized lossy system.** (a) $\varpi$ for HST and ELST are plotted by the red and green lines, respectively. (b) The total optical force for $r = 1.4\ \lambda$ [marked by the black rectangle in (a)] calculated by different stress tensors. The radii for the small cylinders and the entire circular domain are $6 \times 10^{-3}$ $\lambda$ and 1.976 $\lambda$ respectively.

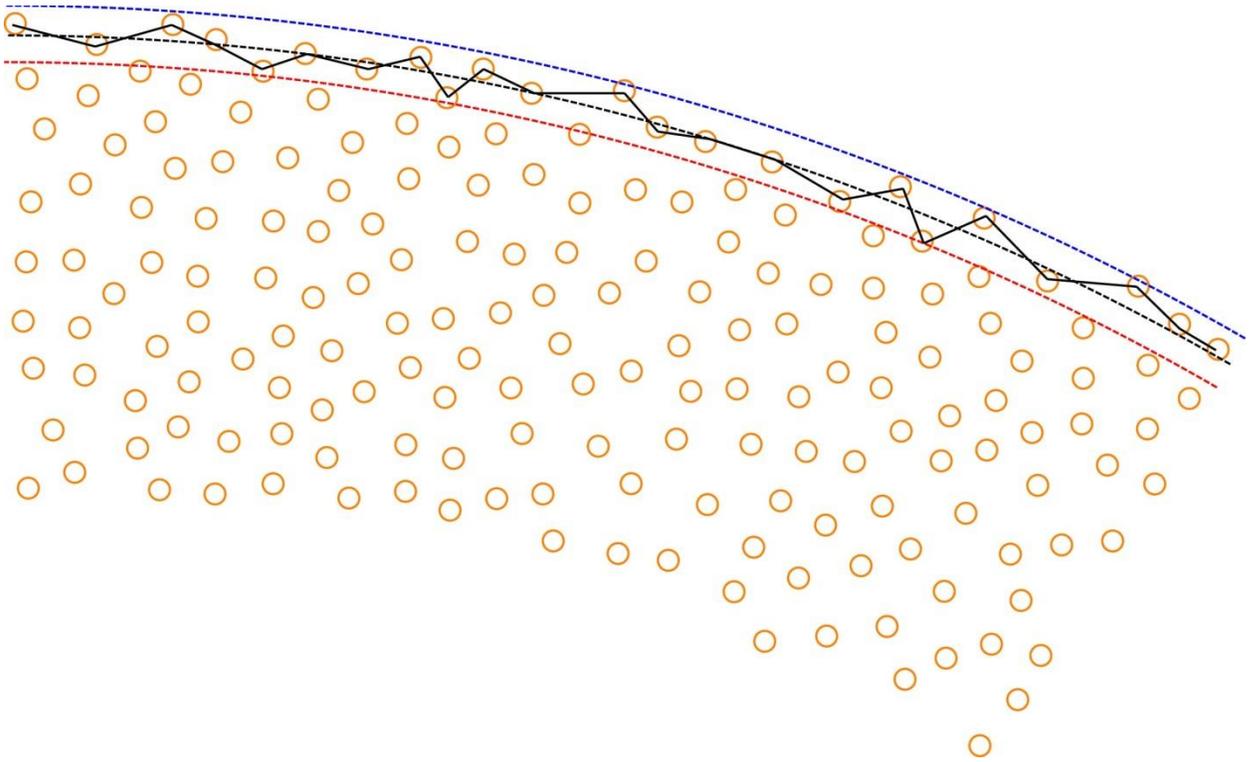

**Figure 9. A portion of a circular piece of the amorphous medium near the boundary.** The orange circles denote randomly distributed cylinders inside a circular region whose boundary is indicated by the blue dashed arc line. The cylinders between the red and blue dashed arc lines are considered as the outermost cylinders. The black solid line connecting all the outermost cylinders forms the actual boundary of the effective medium. We define the the effective medium to be a circle indicated by the black dashed arc line whose radius is determined as the average radial distance between the outermost cylinders and the center of the circular region.